\documentclass[prl,twocolumn,showpacs]{revtex4-1}
\usepackage{amssymb,amsfonts,amsmath} 
\usepackage{graphicx}
\usepackage{color}
\usepackage{hyperref}
\usepackage[final]{pdfpages}

\begin{document}

\title{Renormalization in periodically driven quantum dots}

\author{A.K.\ Eissing}  
\affiliation{Institut f{\"u}r Theorie der Statistischen Physik, RWTH Aachen University 
and JARA---Fundamentals of Future Information
Technology, 52056 Aachen, Germany}
\author{V.\ Meden} 
\affiliation{Institut f{\"u}r Theorie der Statistischen Physik, RWTH Aachen University 
and JARA---Fundamentals of Future Information
Technology, 52056 Aachen, Germany}
\author{D.M.\ Kennes}  
\affiliation{Institut f{\"u}r Theorie der Statistischen Physik, RWTH Aachen University 
and JARA---Fundamentals of Future Information
Technology, 52056 Aachen, Germany}

\begin{abstract} 
We report on strong renormalization encountered in periodically driven interacting quantum dots in the non-adiabatic 
regime. Correlations between lead and dot electrons enhance or suppress the amplitude of driving depending on the sign 
of the interaction. Employing a  newly developed flexible renormalization group based approach for periodic driving 
to an interacting resonant level we show analytically that the magnitude of this effect follows a power law.
Our setup can act as a non-Markovian, single-parameter quantum pump.
\end{abstract}

\date{\today} 
\maketitle

Lately, signal and information processing at the quantum level has attracted sizable interest. 
It can be realized in quantum dot geometries which can be manufactured very precisely, e.g., in 
semi-conductor heterostructures. These provide the basic 
building blocks of future nanoelectronic and quantum information devices \cite{Hansonreview07}.
An experimentally routinely implemented non-equilibrium dot setup is the one of 
periodically varying external fields \cite{Switkes99, Roche2013}. 
Utilizing as an additional parameter the frequency $\Omega$ of the field  one can achieve novel control 
of the quantum dot's transport properties. Examples of such include  a vanishing of certain tunneling 
amplitudes of the dot structure, which is analogous to the coherent destruction in two-level 
systems \cite{Grossmann1991, Grossmann1992}, and pumping of an integer number $n$ of the elementary charge $e$ 
during one cycle \cite{Thouless1983, Pothier1992, Kaestner2014}. The latter is important 
for metrology, where the quantized charge $n e$ could replace the currently used electric current 
standards \cite{Pekolareview13}. 

In the adiabatic limit of a small rate of change of the external field the latter can be used as a small parameter 
in analytic calculations. This was employed to gain valuable insights into the physics of periodically driven 
quantum dots \cite{Brouwer98, Splettstoesser2005, Splettstoesser2006aa, Hiltscher2010, Hernandez2009, Haupt2013}.
The regime in which $\Omega$ is of the order of the (average) tunneling rate $\Gamma$ or larger, however, was 
treated less extensively \cite{Croy2012, Kashuba2012, Moskalets2002}, but 
showed interesting physics, e.g. one parameter pumping \cite{Kaestner2008, Cavaliere2009}. 
In particular, describing the many-electron interplay encountered in non-adiabatically driven quantum dots 
in presence of correlations remains a formidable challenge \cite{Citro2003,Kashuba2012,Hettler1995}. 
We here report on progress on this.

\begin{figure}[t!]
\includegraphics[width=0.85\linewidth,clip]{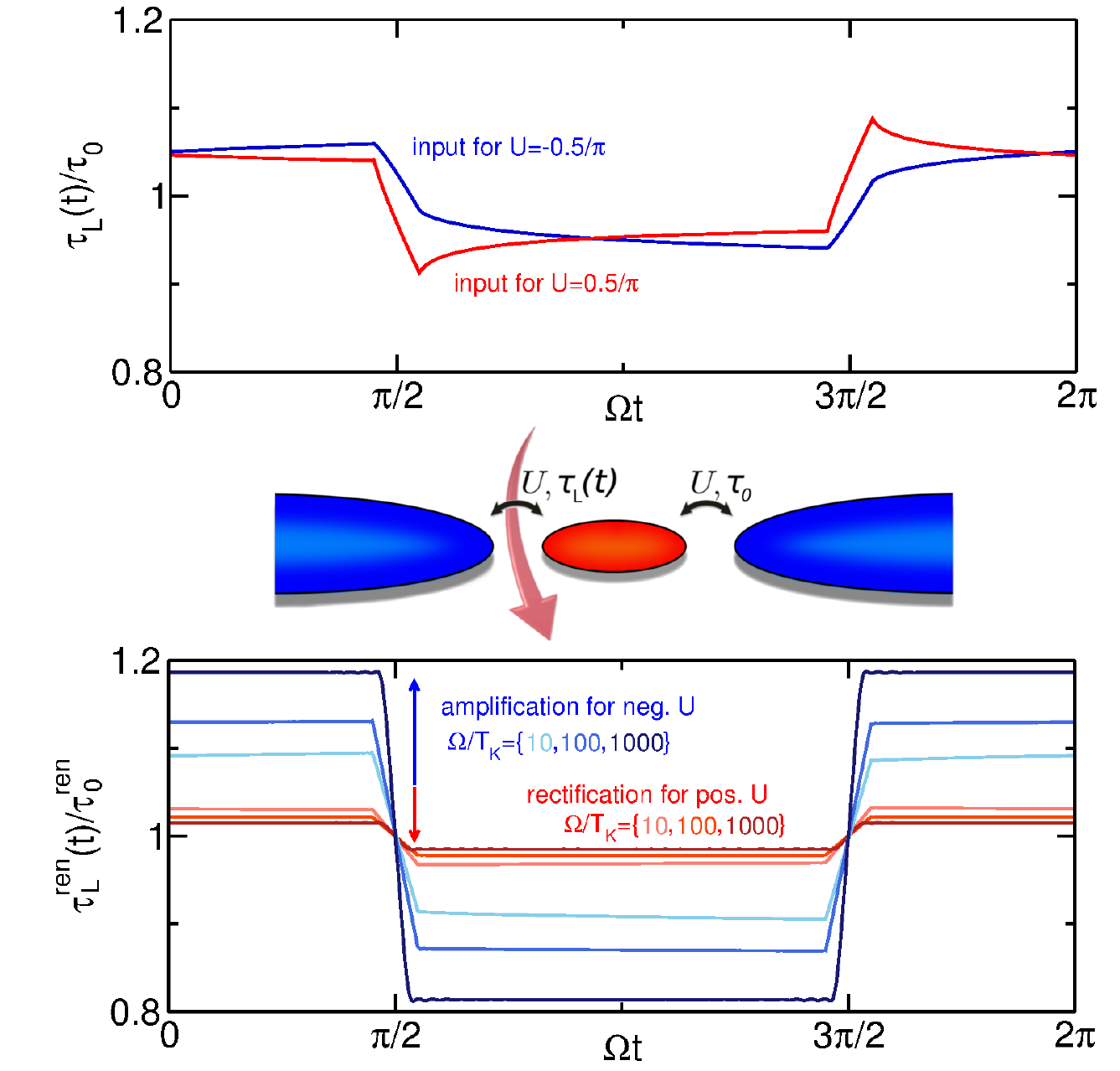}
\caption{(Color online) A periodically varied tunneling amplitude between the left reservoir and the quantum 
dot $\tau_{\rm L}(t)$ is strongly renormalized by an interaction $U$.  The upper panel shows $\tau_{\rm L}(t)$ for one period, which is designed 
to yield a renormalized hopping amplitude of rectangular 
line shape (see the lower panel). It turns out to dependent on $\mbox{sgn}(U)$. The central panel depicts the studied model, which consists of a quantum dot coupled to 
two electronic reservoirs via hopping amplitude $\tau$ and a dimensionless interaction $U$. For $U= \pm 0.5/\pi$ the emergent 
universal energy scale is $T_{\rm K} (U\pi = -0.5) /D = 4.41\cdot 10^{-7}$ and $T_{\rm K} (U\pi = 0.5)/D =2.865 \cdot 10^{-4}$. The Fourier
components of the hopping are $|\tau_{\text{L},k}/\tau_0| = 
0.1 \pi k^{(-1+U)}$. As shown in the lower panel the driving is rectified for positive $U$ (red lines), while it is amplified for 
negative $U$ (blue lines), where both effects enhance with increasing driving frequency $\Omega$ following a power law. }
\label{fig:fig1}
\end{figure}


Following the usual procedure 
\cite{Brouwer98, Splettstoesser2005, Splettstoesser2006aa, Hiltscher2010, Hernandez2009, Haupt2013,Cavaliere2009, Croy2012, Kashuba2012, Moskalets2002, Kohler2005}, 
we model the influence of the external time-periodic fields by time-periodic parameters $p(t)$ of the quantum 
dot setup. In general, all Hamiltonian parameters, such as the dots level positions and the tunneling amplitudes might 
acquire a time dependency. Given these one can compute the relevant observables, such as the current through the 
quantum dot, e.g. using scattering theory \cite{Brouwer98, Moskalets2002}. We emphasize that 
depending on the setup under consideration obtaining closed analytical expressions for observables of 
this time-dependent single-particle problem might turn out to be rather involved (for examples relevant here see below 
and \cite{SM}, \cite{Gomez2013}). In the presence of two-particle interactions a many-particle method must be employed. We use a
newly developed extension of the functional renormalization group (FRG) \cite{Metzner2012} 
to periodically driven systems applicable at arbitrary driving frequency. We show that the emergent correlations 
lead to a strong, qualitative renormalization 
of the time-periodic parameters $p(t)\to p^{\rm ren }(t)$ in the anti-adiabatic regime 
$\Omega \gg \Gamma$.  This affects not only their 
amplitude but their entire line shape (see Fig.~\ref{fig:fig1}). 
For typical quantum dot models the renormalization 
is not accessible by perturbation theory in either the two-particle interaction or the tunnel coupling. It can be 
exploited to flexibly design the effective driving and thus the time periodic current by controlling the 
frequency (see Fig.~\ref{fig:fig1}). Within our truncation 
of the hierarchy of FRG flow equations the $ p^{\rm ren }(t)$ can directly be plugged into the non-interacting 
expressions for observables. Following this two step procedure of (i) setting up and solving RG equations for the 
time-dependent parameters and (ii) substituting them into noninteracting expressions we device a method
which allows one to obtain numerical results for arbitrary dot parameters and analytical ones in limiting
cases such as e.g.\,the anti-adiabatic regime of interest here. 
    
To depict the parameter renormalization in a transparent fashion, we concentrate on a spin polarized, single level 
quantum dot. It hosts a fermionic degree of freedom and is coupled to two 
infinite, equilibrated reservoirs. We 
consider the left hopping $\tau_{\rm L}(t)=\sum_{k} \tau_{{\rm L},k} e^{i\Omega k t}$ to be time-periodic, allowing 
for an arbitrary line shape, i.e.\,arbitrary Fourier 
amplitudes $\tau_{{\rm L},k}$. The right hopping is set constant $\tau_{\rm R}(t)=\tau_{{\rm L},0}\equiv \tau_{0} $, 
where we focus on left-right symmetric mean values. 
Additionally, we consider a capacitive density-density type interaction $U$ between electrons residing 
in the reservoirs and on the dot as well as an onsite energy $\epsilon$, which for simplicity are both 
assumed to be time-independent.
The underlying model is known as the interacting resonant level model 
(IRLM) \cite{Schlottmann1980a, Schlottmann1982a, Doyon2007, Borda2007, Boulat2008, Karrasch2010, Karrasch2010b}, 
which is the standard model for correlated quantum dots dominated by charge fluctuations. 
The Hamiltonian is $H=H_{\rm dot}+H_{\rm coup}+H_{\rm res}$,
with $H_{\rm dot}= \epsilon d^\dagger d$  and
\begin{align}
H_{\rm coup}=& \sum_{q,\alpha={\rm L,R}}\tau_{\alpha}(t)\left(d^\dagger c_{q,\alpha}+{\rm h.c.}\right)\notag\\
&+\left(d^\dagger d -\frac12\right)\sum_{q,q',\alpha={\rm L,R}}\frac{U_{\alpha}}{\rho_{\alpha}^{(0)}} :c_{q,\alpha}^\dagger c_{q',\alpha}: .
\end{align}
We use second quantization notation, where the dot and reservoir operators are $d$ and $c$, respectively, 
and $:...:$ denotes normal ordering.
The reservoirs given by $H_{\rm res}$ only enter via their local density of states $\rho_\alpha(\omega)$. 
For convenience we choose a symmetric Lorentzian $\rho_\alpha(\omega)=\rho_{\alpha}^{(0)}\frac{D^2}{D^2+\omega^2}$ with 
$\rho_{\alpha}^{(0)} = \frac{1}{\pi D}$, 
where the bandwidth $D\gg |\tau_{\alpha}(t)|, |\epsilon|$. In this (scaling) limit the details 
of the energy dependence of $\rho_\alpha$ are irrelevant.
We assume $U_\alpha=U$ but consider both cases of positive and negative 
interaction. The latter might effectively be realized for a quantum dot coupled to phonons, with the phonon frequency  in 
the adiabatic limit \cite{Eidelstein2013}. 

The IRLM can be mapped to the anisotropic Kondo model \cite{Heyl2010} and our results 
are of relevance for this model as well. The Kondo effect under periodic driving was 
studied earlier \cite{Hettler1995,Kaminski2000,Ng1996} employing the single-impurity Anderson and 
the isotropic Kondo model.

We show that in anti-adiabatic driving the correlations induced even by a small $U$, strongly renormalize the 
Fourier coefficients   
$\tau_{{\rm L},k}\to \tau^{\rm ren}_{{\rm L},k}$. This is captured by the central analytical result of our Letter \cite{SM}
\begin{align}
\tau^{\rm ren}_{{\rm L},0}&\sim  f({\tau_{0},U) } \tau_0 , \label{eq:threnk0} \\
\tau^{{\rm ren}}_{{\rm L},k}&\sim  \Omega^{-U }  (ik)^{-U }\tau_{{\rm L}, k} , \;\;  k \neq 0 \label{eq:threnkn0}
\end{align} 
where $\Omega$ is the largest scale (besides the bandwidth $D$). 
We furthermore assume a small amplitude of driving $|\tau^{\rm }_{{\rm L},k\neq 0}/\tau^{\rm }_{0}|\ll 1$. 
Note, that the formalism we device can also be used beyond these limits and for more general setups with all dot parameters 
being time-periodic. 
While $\tau^{\rm ren}_{{\rm L},k=0}$ Eq.~\eqref{eq:threnk0} is independent of $\Omega$,  the higher harmonics 
 Eq.~\eqref{eq:threnkn0} are suppressed or amplified in a power-law fashion $\Omega^{-U}$ with increasing driving frequency, depending on the sign of the 
interaction $U$. The factor $k^{-U}$ leads to a change of the line shape of $\tau^{\rm ren}_{{\rm L}}(t)$ 
after transforming back to time space. Consequently, varying $\Omega$ for a fixed $U$, one finds a constant 
mean value and equal line shape of the renormalized $\tau^{\rm ren}_{{\rm L}}(t)$, but a strongly increased or suppressed amplitude.  
This renormalization can be exploited as an amplifier or rectifier.
Let us assume that one is interested in achieving a given line shape of the hopping amplitude 
$\tau^{\rm ren}_{{\rm L}}(t)$ and thus a given shape of the time-periodic current. We here choose the former to be 
rectangular, but any form could be targeted. At a given value of $U$ we design the external driving such that 
$(ik)^{-U}\tau^{\rm }_{{\rm L},k}$ is equal to the Fourier coefficients of a periodic rectangular pulse. 
This is depicted in Fig.~\ref{fig:fig1} for  $U\gtrless 0$ at $\epsilon=0$. Here we already 
use the emergent scale $T_{\rm K}$ as the unit of energy, which will be introduced next. 

\noindent {\it Power laws in the IRLM}---Already in equilibrium observables of the 
IRLM, e.g. the charge susceptibility
\begin{equation}
\chi=\left.\frac{dn}{d\epsilon}\right|_{\epsilon=0}\sim \left(\frac{\tau_{\alpha}}{D}\right)^{\nu(U)},
\end{equation}   
exhibit power-law scaling with interaction dependent exponents. Plain perturbation theory (in $U$) 
thus leads to logarithmic terms. Various approaches were devised to resum 
these \cite{Schlottmann1980a, Schlottmann1982a, Doyon2007, Borda2007, Boulat2008,Karrasch2010}. 
The susceptibility can be used to define an emergent low-energy scale \cite{Schlottmann1982a},
$T_{\rm K}=-2/(\pi \chi)$.
Universal power laws were encountered in the (static) bias voltage driven non-equilibrium steady state, e.g. in 
the current-voltage characteristics \cite{Doyon2007,Boulat2008,Karrasch2010, Karrasch2010b, Kennes2013a}. 
Furthermore, even the transient relaxation dynamics shows power laws 
in the time variable with interaction dependent exponent \cite{Karrasch2010b, Kennes2012b, Kashuba2013a}. This renders the dynamics 
of the periodically driven IRLM a likely candidate for power-law renormalization.  

\noindent {\it Renormalization group approach to periodically driven quantum dots}---Here we  introduce a reformulation of the general time-dependent FRG of Ref.~\cite{Kennes2012b} 
to the time-periodic problem of interest by making use of the Floquet theorem. This allows one to efficiently treat 
time-periodic setups with arbitrary $\Omega$ and gain analytical insights in limiting cases, which are difficult to obtain otherwise.
Within the FRG framework a suitable chosen cutoff $\Lambda$ is introduced. The quantum many-body 
problem is rephrased in terms of a coupled set of infinitely many 
differential flow equations with respect to $\Lambda$ for the one-particle irreducible 
(Keldysh) vertex functions \cite{Metzner2012}. The only approximation consists in truncating this set, 
here by suppressing the renormalization of the two-particle vertex (effective interaction), which is $\mathcal{O}(U^2)$. As a 
consequence, only the self-energy acquires a RG flow. The effect of the interaction $U$ is incorporated successively 
from high to low  energies, resulting in renormalized single-particle parameters (self-energy). 
  
We are interested in the periodically driven steady state and thus assume that the driving as well as the dot-reservoir-coupling 
were switched on in the distant past. As usual in non-equilibrium retarded, advanced and Keldysh Green's functions 
are introduced and written in standard matrix form \cite{Larkin1975}
\begin{equation}
\hat{G} = 
\begin{pmatrix}
  G^{\text{ret}} & G^K  \\
  0 &  G^{\text{adv}} \\
 \end{pmatrix}.
\end{equation}
An analogous structure is found for the self-energy.
We employ the auxiliary infinite-temperature reservoir cutoff described in \cite{Kennes2012}, to introduce the flow 
parameter $\Lambda$ in every non-interacting Green's function.
The two-times Green's functions are determined by the Dyson-equation
\begin{equation}
\Big [i \frac{\partial}{\partial t}- \hat{\epsilon}_{\text{eff}}(t)\Big ] \hat{G} (t,t') = \sigma_z \delta (t,t'),
\end{equation}
where the single-particle part of the Hamiltonian $\hat{\epsilon}_{\text{eff}}(t)$ incorporates the effect of 
the self-energy; the latter remains time local in our approximation \cite{Kennes2012}.
This differential equation can  be solved for  time-periodic $\hat{\epsilon}_{\text{eff}}(t)$ exploiting the Floquet 
theorem \cite{Floquet1883}. One writes 
\begin{equation}
 G^X(t,\omega)=\int dt' e^{i\omega (t-t')}G^X(t,t'),
 \end{equation}
and 
\begin{equation}
G_k^X(\omega) = \frac{|\Omega|}{2\pi} \int dt\, e^{i k\Omega t} G^X(t,\omega),
\end{equation}
with $X\in \{{\rm ret,K,adv}\}$, $k \in {\mathbb Z}$  being a discrete Floquet index, and $\omega$ a continuous real 
frequency \cite{Wu2008, Wingreen1993, Tsuji2008}. 
The self-energy and observables of interest are transformed accordingly. Single-particle-like indices are easily 
incorporated in a FRG framework increasing the complexity only in a polynomial fashion \cite{Rentrop2014}. 
Formulated in this way the FRG approach to the steady state of the time-periodic problem is closely related 
to the static non-equilibrium FRG \cite{Metzner2012}. The technical details of the Floquet FRG including the 
full RG flow equations for arbitrary (amplitude and line shape) time-periodic dot parameters will be presented 
elsewhere \cite{longFFRG2015}. For general parameters the flow equations must be solved numerically with the 
only additional challenge compared to established equilibrium and static non-equilibrium FRG schemes that a 
sufficient number of higher harmonics $k$ must be kept; this can be chosen to be $\mathcal{O}(1000)$ without 
exceeding available numerical resources. We note that all our figures show the results of the numerical solution 
of the full set of flow equations. 

In the limit of small amplitudes $\tau_{\alpha,k}$ one can derive simplified RG equations which for large $\Omega$ 
can then be solved analytically. This allows us to gain valuable analytical insights into the mechanism of power-law
suppression/amplification described above. For simplicity we concentrate on $\epsilon=0$. 
In this case the flow equations of the hopping amplitudes $\tau_{\alpha,k}^{\Lambda}$ are the only relevant ones, while the 
renormalization of the onsite energy $\epsilon^{\Lambda}_k$ is zero (or $\mathcal{O}(U^2)$ for $\epsilon\neq 0$) \cite{Karrasch2010}. 
To leading order in $\tau_{\alpha,k}^{\Lambda}$ and $U$ the differential flow equations for 
$\partial_{\Lambda}\tau^{\Lambda}_{\alpha,k}$ decouple and with $\tau^{\Lambda}_{\text{L},k=0} = \tau^{\Lambda}_{\text{R},k=0} 
\equiv \tau^{\Lambda}_{0}$ are
 \begin{equation}
\partial_{\Lambda} \tau^{\Lambda}_{\text{L},k\neq 0}= - U  \frac{ \tau^{\Lambda}_{\text{L},k} \Lambda/D^2}{\frac{\Lambda^2}{D^2}+ (4 \frac{|\tau_0|^2}{D^2}+ \frac{ik\Omega}{D}) \frac{\Lambda}{D}+ \frac{2i|\tau_0|^2 k \Omega}{D^3}+\frac{4|\tau_0|^4}{D^4}}
 \end{equation}
and
 \begin{equation}
\partial_{\Lambda} \tau^{\Lambda}_{0}= - U \frac{\tau^{\Lambda}_{0}/D}{(\Lambda/D)^2 + \Lambda/D + 2(\tau_{0}/D)^2}.
 \end{equation}
As emphasized above the flow of $\tau^{\Lambda}_{0}$ is independent of $\Omega$ while in the one of 
$\tau^{\Lambda}_{\text{L},k\neq 0}$, $k \Omega$ acts as a cutoff.  One can solve these 
equations in a closed form (see \cite{SM}) yielding Eqs.~\eqref{eq:threnkn0} and \eqref{eq:threnk0} for large $\Omega$. The consequences 
of this renormalization as well as its potential as an amplifier or rectifier, were discussed above. 
 
 \begin{figure}[t]
\includegraphics[width=.9\linewidth,clip]{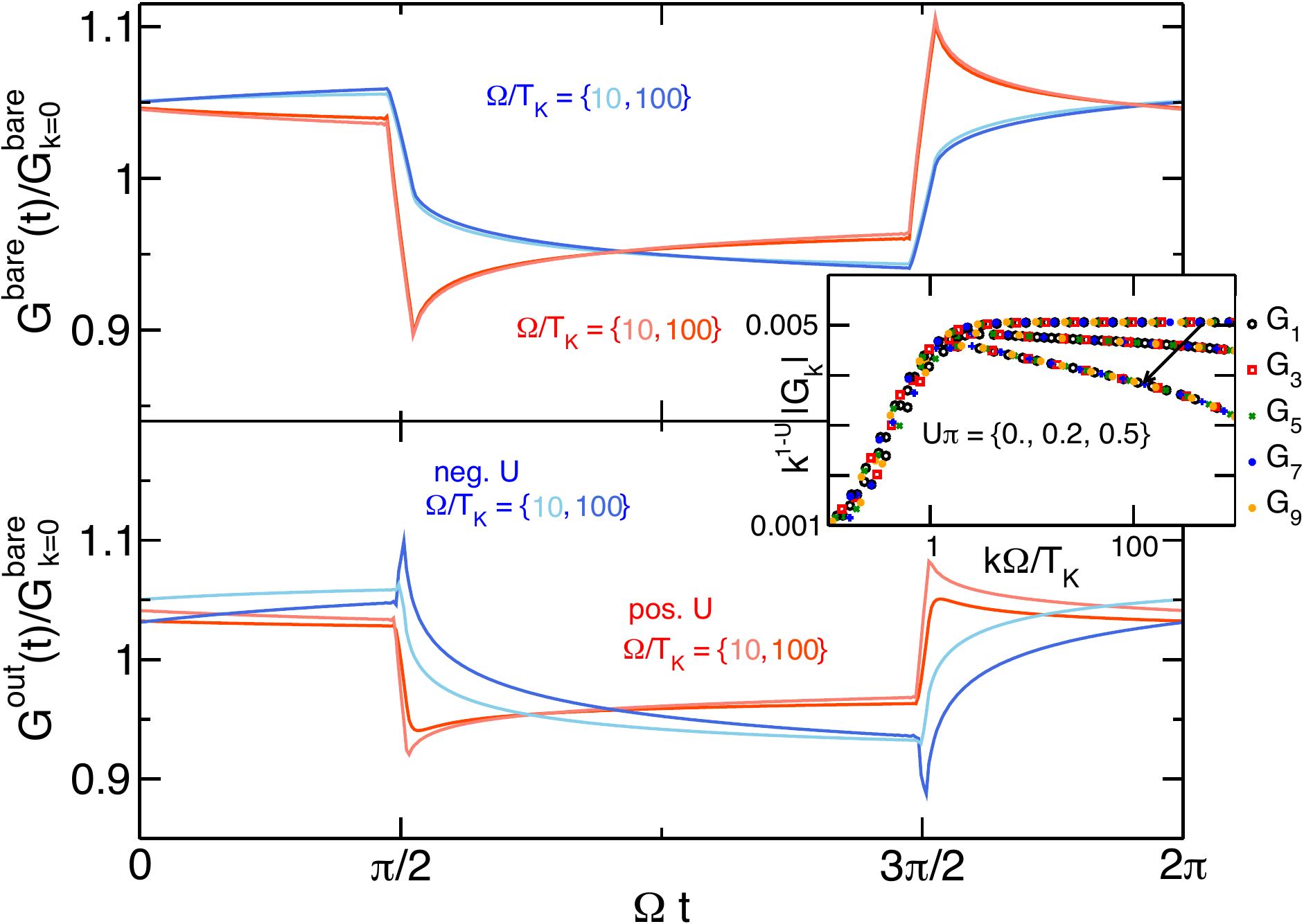}
\caption{(Color online) The linear conductance $G_{\rm L}(t)$ for the non-interacting case (upper panel) as well as renormalized 
by the interaction (lower panel) for two values of driving frequency $\Omega$ for the parameters of Fig.\ref{fig:fig1}. 
The amplification/rectification effect of the hopping is reflected.  The inset shows that the (odd) higher harmonics 
(even ones are zero) of the conductance $G_{\text{L},k}$ collapse for different $k$ if  $G_{\text{L},k}$ and the driving frequency are scaled by a factor of $k^{1-U}$ and $k$, respectively.}
\label{fig:fig2}
\end{figure}

\noindent {\it Observables}---In our truncation of the exact hierarchy of flow equations one can 
substitute the renormalized time-periodic parameters into the 
non-interacting expressions of observables, e.g. the current through the quantum dot structure or the 
occupancy of the dot. 
This allows one to obtain (approximate) closed form expressions in the presence of interactions and is a particular advantage of our 
approximate RG formalism. 
We here focus on the time-periodic current  $J_\alpha(t)=\sum_{k} J_{\alpha,k} e^{i k \Omega  t}$ leaving reservoir $\alpha$ which 
is given by
\begin{align}
& J_{\alpha,k}= \,\frac{1}{4\pi} \sum_{k'} \int d\omega\,\Big[\Sigma^{\text{ret}}_{\alpha,-k-k'} (\omega+k'\Omega) \,G^K_{k'} (\omega) 
\nonumber \\  & - G^{\text{ret}}_{-k-k'} (\omega+ k'\Omega) \,\Sigma^K_{\alpha,k'} (\omega) \Big] +  [-k \rightarrow k]^*
\label{eq:cur}
\end{align}
where the mean value is the zeroth component defined as $J_{\alpha,k=0} = \frac{1}{T} \int_0^T J_\alpha(t)$ and $\Sigma^{X}_{\alpha,k}$ 
denotes the self-energy of the reservoir $\alpha$ \cite{Wingreen1993}. The components of the Green's function and self-energy 
contain the interaction via the renormalized parameters. We exemplify the impact of the parameter renormalization 
of the periodically driven dot on the current for two examples.

\noindent {\it Linear response current under the driving of Fig.~\ref{fig:fig1}}---We first consider the driving protocol 
of Fig.~\ref{fig:fig1} and study the linear response of $J(t)$  to an additional small static bias voltage 
$V=\mu_{\rm L}-\mu_{\rm R}$ between the left and right reservoir, where $V$ is much smaller than any other scale.
The effective single-particle problem of a dot with a time-periodic left hopping of rectangular shape 
(see Fig.~\ref{fig:fig1}) is rather involved as the different Floquet `channels' indicated by $k$ are all coupled.
We thus refrain from further simplifying Eq.~\eqref{eq:cur}. 
We define $G_\alpha(t)=\lim_{V \to 0} dJ_\alpha(t)/dV$ as the linear conductance and analogously its Fourier coefficients 
$G_{\alpha,k}=\lim_{V \to 0} dJ_{\alpha,k}/dV$. The upper panel of Fig \ref{fig:fig2} shows the bare, i.e.\,non-interacting 
$G_{\rm L}(t)$ resulting from the external driving of Fig.~\ref{fig:fig1} for $\Omega/T_{\rm K}=10$ and $100$. 
As seen in the lower panel the overall amplitude of the renormalized conductance is also rectified/amplified 
as expected from Eq.~\eqref{eq:threnkn0}. 
 
The inset of Fig.~\ref{fig:fig2} depicts the (odd) higher harmonics of the conductance
(even ones are zero) for different $U$. 
A collapse of the conductance for different $k$ is found at fixed $U$ when rescaling $G_{\text{L},k}$ 
by $k^{1-U}$ and $\Omega$ by $k$. This shows that although $G_{\rm L}(t)$ depends on the renormalized
dot parameters in an involved way, universal scaling established in the IRLM in and out-of-equilibrium 
is also realized for periodic driving. 

 \begin{figure}[t]
\includegraphics[width=.9\linewidth,clip]{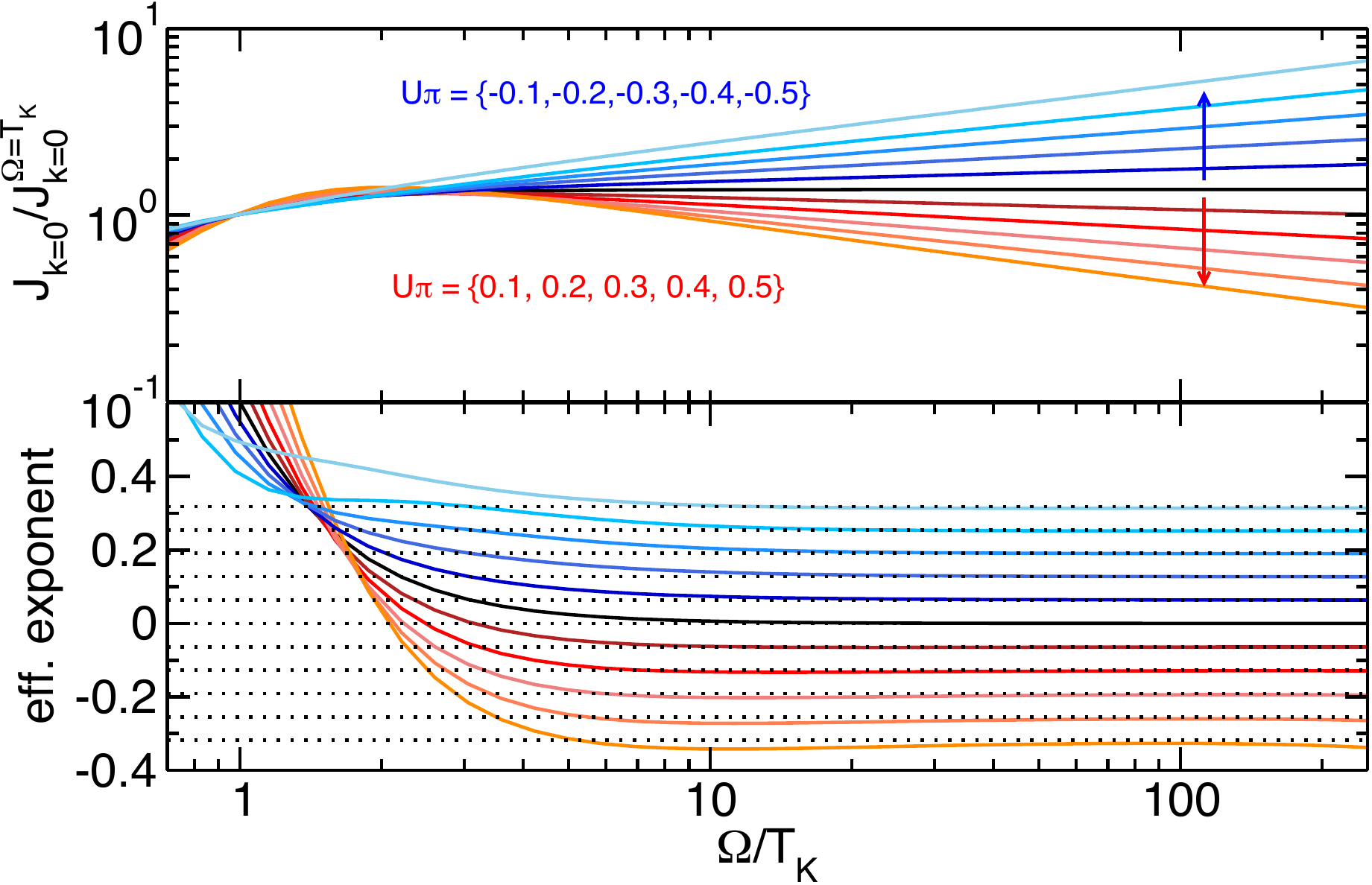}
\caption{(Color online) The upper panel shows the pumping power $J_{\text{L},k=0}(\Omega)$ in units of $J_{\text{L},k=0}(\Omega=T_K)$ for a single parameter pump. $\tau_{\text{L}} = \tau_{0} + \Delta \tau \sin (\Omega t)$ with and 
$\frac{\Delta \tau}{\tau_{0}} =0.05$. It shows again an amplification for negative as well as a rectification 
for positive values of $U$ and inherits the power-law behavior of the hopping amplitude. The lower panel 
shows the logarithmic derivative of $J_{\text{L},k=0}$ indicating the effective exponent $\alpha$ 
of the power law behavior, which is in a very good agreement with the predicted exponent (dotted lines). The $T_{\rm K}/D$ are \{$4.41\cdot 10^{-7},8.4 \cdot 10^{-7}, 2.2\cdot 10^{-6}, 5.55 \cdot 10^{-6}, 1.24 \cdot 10^{-5},$ $2.5 \cdot 10^{-5}, 4.62\cdot 10^{-5}, 8.43 \cdot 10^{-5}, 1.28\cdot 10^{-4}, 1.96\cdot 10^{-4}, 2.865\cdot 10^{-4}$\} 
for $U\pi = \{-0.5, -0.4,..., 0., ...,0.4,0.5\}$. }
\label{fig:fig3}
\end{figure}

\noindent {\it Pumping}---Finally, we consider the quantum dot as a single-parameter pump. In quantum pumping, one usually studies 
a sinusoidal time-dependence of the dot parameter(s) \cite{Brouwer98, Splettstoesser2005, Splettstoesser2006aa, Hiltscher2010, Hernandez2009, Haupt2013}. 
Assuming $\tau_{\rm L} (t)=\tau_{0} + \Delta \tau \sin (\Omega t)$ we follow this convention. 
By breaking the particle-hole symmetry via (arbitrarily) choosing $\epsilon=0.4 T_{\rm K}$, $J_{\text{L},k=0} \neq 0$ 
for non-adiabatic driving.  $J_{\text{L},k=0}$ describes the average amount of pumped charge per unit time 
and thus indicates the pumping power. It can be related to the pumped charge per cycle
$Q= T J_{\text{L},k=0}$ \cite{Brouwer98, Moskalets2002, Splettstoesser2005}. 
Following our two step approach we derived \cite{SM} a closed form expression for 
$J_{\text{L},k=0}$ based on Eq.~\eqref{eq:cur} (see also \cite{Gomez2013}) -- which is straightforward as the Floquet channels
are only `weakly' coupled for sinusoidal driving -- and substitute the renormalized parameters, 
which yields to order $(|\tau^{{\rm ren}}_{{\rm L},1}| / \tau^{\rm ren}_{{\rm L},0})^2$ and for large $\Omega$
 \begin{equation}
 J_{\text{L},k=0}\stackrel{\Omega\gg\epsilon}{=}\frac{1}{2\pi}\left(\frac{|\tau^{{\rm ren}}_{{\rm L},1|}}{\tau^{\rm ren}_{{\rm L},0} }\right)^2 
T_K\arctan\left(\frac{2\epsilon}{T_K}\right)
.
 \end{equation}
In the upper panel of Fig.~\ref{fig:fig3} we show $J_{\text{L},k=0}$ for different $U$. 
A clear power law $J_{\text{L},k=0}\sim \Omega^{\gamma}$ with $\gamma=-2U$ manifests in the pumped charge at large driving 
frequency, which is a direct consequence of the renormalization of the tunneling amplitude $\tau^{{\rm ren}}_{{\rm L},1}$ 
Eq.~\eqref{eq:threnkn0}. Our main result of non-perturbative power-law amplification or rectification can thus be 
observed in the straightforwardly measurable pumping power. 
We note that a numerical solution of the full set of FRG flow equations is sufficiently accurate  
to extract the exponent $\gamma$ via a logarithmic derivative $d\ln(J_{\text{L},k=0})/d\ln(\Omega)$,
(implemented as centered differences) which is a numerically very sensitive measure; see the 
lower panel of Fig.~\ref{fig:fig3}. This exemplifies that from the data for observables detailed information can be extracted 
which is crucial in cases in which the effective scattering problem is to involved to gain analytical
insights.

The pumped current is a consequence of the non-Markovian 
nature of the reservoirs and it is not obvious how to capture it in standard approaches relying on a tunneling rate which 
depends on a single time argument. In contrast to the quantum pump described in 
Ref.~\cite{Cavaliere2009} (realized for $\Omega \lessapprox \Gamma$) it cannot be understood in terms of interaction 
induced effective phase differences of the dot parameters.

\noindent {\it Conclusion}---We have shown that in quantum dots, where one hopping amplitude between the quantum dot and the electronic reservoir 
is varied periodically, the effective hopping amplitude is renormalized strongly by correlations, resulting in power-law scaling of 
the Fourier coefficients of the hopping. This is reflected in the line shape and the overall amplitude of the renormalized hopping. 
Depending on the sign of the interaction a power-law amplification or rectification is achieved. This renormalization in turn manifests in observables such as the linear conductance in the presence of an additional infinitesimal static bias voltage or the 
pumped charge of a single parameter quantum pump. The particular setup of only one hopping amplitude chosen as time-periodic can 
be easily generalized to one with any one (as e.g.\,$\epsilon (t)$) or even more parameters being time-periodic within our newly developed 
RG approach. The renormalization effects are relevant to experiments, as they dominate for driving frequencies of the order of 
$10 T_{\rm K}$, which for dot geometries with an effective hybridization of $1$ MHz is within an accessible regime.

\noindent \textit{Acknowledgments}---We thank J. Splettst\"o{\ss}er for discussions. This work was supported  by the DFG (RTG 1995).

\bibliographystyle{apsrev4-1}
\bibliography{FloquetBib}

%
%
%
%
%
%
%

\clearpage
\appendix


\section*{Supplementary material}

# Renormalization in periodically driven quantum dots
## Supplementary Material

Here we provide (a) the analytical solution of the flow Eqs. (9) and (10) as well as (b) the derivation of the closed expression of the pumping power of the effective non-interacting model given in Eq. (12) of the main text. This exemplifies how we are able to gain analytic insights into the renormalization effects employing our functional renormalization group scheme. After the flow equations have been derived these can be solved analytically (in certain limits). Due to the truncation used here it suffices to plug the renormalized parameters obtained from the flow equations into the non-interacting expressions of the observables of interest.

### (a) Analytical solution of the flow equations

The differential Eqs. (9) and (10) can be solved analytically, yielding ($\tau_0^{\Lambda=\infty} = \tau_0$)

$$
\begin{aligned}
\frac{\tau_0^{\Lambda=0}}{\tau_0} &= \left[ \frac{1 - \sqrt{1 - 8(\tau_0/D)^2}}{1 + \sqrt{1 - 8(\tau_0/D)^2}} \right]^{-U[1-8(\tau_0/D)^2]^{-1/2}} \\
&\overset{D \gg \tau_0}{=} \left( \frac{2\tau_0^2}{D^2} \right)^{-U}
\end{aligned}
\tag{S1}
$$

and

$$
\begin{aligned}
\frac{\tau_{L,k\neq0}^{\Lambda=0}}{\tau_{L,k\neq0}^{\Lambda=\infty}} &= e^{\frac{U}{2k\Omega}\left(2ik\Omega + \frac{4|\tau_0|^2}{D}\right)\arctan\left(\frac{k\Omega}{D+2|\tau_0|^2/D}\right)} \left( D + \frac{2|\tau_0|^2}{D} \right)^{\frac{2iU|\tau_0|^2}{D(k\Omega)}} \left[ k^2\Omega^2 + \left( D + \frac{2|\tau_0|^2}{D} \right)^2 \right]^{\frac{U(k\Omega - 2i|\tau_0|^2/D)}{2k\Omega}} \\
&\quad \times e^{\frac{-U}{2k\Omega}\left(2ik\Omega + 4\frac{|\tau_0|^2}{D}\right)\arctan\left(\frac{k\Omega}{2|\tau_0|^2/D}\right)} \left( \frac{2|\tau_0|^2}{D} \right)^{\frac{-2iU|\tau_0|^2}{Dk\Omega}} \left[ k^2\Omega^2 + \left( \frac{2|\tau_0|^2}{D} \right)^2 \right]^{\frac{-U(k\Omega - 2i|\tau_0|^2/D)}{2k\Omega}} \\
&\overset{D \gg \tau_k}{=} e^{\frac{U}{2k\Omega} \frac{4|\tau_0|^2}{D} \arctan\left(\frac{k\Omega}{2|\tau_0|^2/D}\right)} \left[ \frac{k^2\Omega^2 + 4(|\tau_0|^2/D)^2}{D^2} \right]^{-U/2} e^{-iU\arctan\left(\frac{k\Omega}{2|\tau_0|^2/D}\right)} \left[ \frac{4(|\tau_0|^2/D)^2}{k^2\Omega^2 + 4(|\tau_0|^2/D)^2} \right]^{\frac{U|\tau_0|^2/D}{k\Omega}} \\
&\overset{\Omega \gg T_K}{\longrightarrow} \left( \frac{k\Omega}{D} \right)^{-U} (i)^{-\text{sign}(k)U}
\end{aligned}
\tag{S2}
$$

In the last step, we have specified that the frequency $\Omega \gg T_K$ which yields Eq. (3) of the main text (disregarding the irrelevant proportionality $\sim D^{-U}$).

### (b) Single parameter quantum pump

It is a particular advantage of our functional renormalization group scheme that once the flow equations have been solved observables can be calculated straightforwardly by considering an effectively non-interacting system with renormalized parameters. Those renormalized parameters include (via the RG procedure) the influence of the two-particle interaction. This allows for a derivation of analytic expressions for the observables, which depend on the details of the driving protocol. Here we consider as an example $\tau_L(t) = \tau_0 + \Delta\tau \sin(\Omega t)$ and finite $\epsilon$, which includes the quantum pump described in the main text. We aim at a closed form expression for the pumping power $J_{L,k=0}$ to order $\mathcal{O}(\tau_{L,1}^2)$ in the high frequency regime. Given the renormalization of $\tau_0$ and $\tau_{L,1}$ (see Eqs. (S1) and (S2) and Eqs. (2) and (3) of the main text) this way provide analytical insights into the power-law renormalization of $J_{L,k=0}$ discussed in the main text.

It is straightforward to solve the scattering problem of non-interacting fermions for the resonant level model with renormalized $\tau_{L,k}$ numerically. However, to reveal insights into the pumping mechanism (and its power-law suppression/amplification) we choose a more instructive path in the following. Remind that the reservoirs are modeled via a Lorentzian density of states $\rho_\alpha(\omega) = \rho_\alpha^{(0)} \frac{D^2}{D^2 + \omega^2}$ with $D$ being the largest scale (scaling limit). This description is

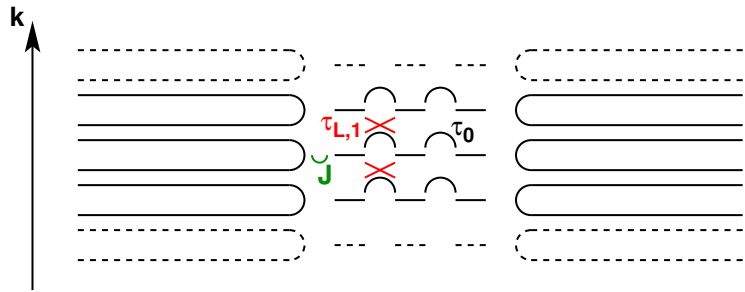

FIG. S1: (Color online) The non-interacting driven quantum dot can be formally mapped to a non-driven one-dimensional system by interpreting the Floquet index $k$ as a spatial one.

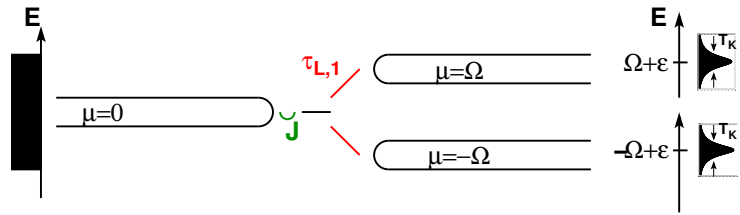

FIG. S2: (Color online) To $\mathcal{O}(\tau_{L,1}^2)$ the pumped current can be evaluated in an effective three-terminal setup. To the left and right the energy dependent density of states of the three reservoirs are shown.

equivalent to keeping one site of the reservoir explicit and coupling it to a structureless reservoir with hybridization $D$. In this language the single reservoir site kept explicitly is coupled to the central level via $\tau_0$. Next we note that including the Floquet index $k$ for the periodically driven system is mathematically identical to enlarging the dimensionality of the system (the index $k$ then takes the role of an additional spatial index, see [25]). Along the infinite '$k$-direction' replicas of the initial system with all energy levels shifted by $k\Omega$, have to be included. The different replicas are coupled via $\tau_{L,k}$ (where the Floquet index $k$ indicates the range of the coupling in the '$k$-direction'). With this construction one can map the non-interacting periodically driven zero-dimensional quantum dot to an one-dimensional system in the steady state (no periodic driving). As we consider a sinusoidal form of $\tau_L(t)$ only $\tau_{L,1}$ is non-zero and couples neighboring replicas. The mapping is sketched in Fig. S1

We aim at an analytical expression for the current $J$ through the central channel $k = 0$ of the mapped model shown in Fig. S1 to $\mathcal{O}(\tau_{L,1}^2)$ in the large frequency $\Omega$ limit. To $\mathcal{O}(\tau_{L,1}^2)$ the influence of the replicas below $k = -1$ and above $k = 1$ (dotted in Fig. S1) can be neglected and the contribution of the $k \neq 0$ channels to the $k = 0$ one can be evaluated in their respective equilibrium (where all $\tau_{L,1}$ are set to zero).

To sum up, the scattering problem through the central channel $k = 0$ is thus dressed by temporary excursions of particles into the $k = \pm 1$ channel (and back). The temporary excursions exhibit an equilibrium background in the $k = \pm 1$ channel. We aim at the current leaving the left reservoir [denoted by $J(= J_{L,k=0})$ in Fig. S1]. Left and right reservoirs are held at the same chemical potential and thus to $\mathcal{O}(\tau_{L,1}^0)$ there is no net current $J_{L,k=0}$. Also temporary excursions of particles from the central site at $k = 0$ to the left site of the $k = \pm 1$ channel and back to the central site of $k = 0$ always respect left-right symmetry and give no net contribution to the current $J_{L,k=0}$ from left to right. Particle excursions from the central site at $k = 0$ to the left site of the $k = \pm 1$ channel and back to left site of $k = 0$ break the left-right symmetry, but are suppressed by $1/\Omega$ and provide no contribution for large frequency. The only relevant process to $\mathcal{O}(\tau_{L,1}^2)$ is thus the one from the left site of $k = 0$ to the central site of $k = \pm 1$ and back. Now we consider the left site of the $k = 0$ channel, which within $k = 0$ is coupled much more strongly to the left ($\sim D$) than to the right ($\sim t_0^2/D$). Therefore the influence of the coupling of the left site to the central site of the $k = 0$ channel will be neglected in leading $\mathcal{O}(1/D)$. Calculating $J_{L,k=0}$ thus effectively maps to the problem of a single site (former left site of $k = 0$) coupled to three reservoirs. This is depicted in Fig. S2.

Using the Landau-Büttiker equation [49] one now finds Eq. (12) of the main text by

$$
\begin{aligned}
J_{\mathrm{L},k=0} &= \frac{1}{2\pi} \int\limits_{-\infty}^{\infty} dE \; [\Theta(-E) - \Theta(-E+\Omega)] \, \frac{2}{D} \frac{|\tau_{L,1}|^2 T_K/2}{(E-\epsilon-\Omega)^2 + (T_K/2)^2} \\
&\quad + \frac{1}{2\pi} \int\limits_{-\infty}^{\infty} dE \; [\Theta(-E) - \Theta(-E-\Omega)] \, \frac{2}{D} \frac{|\tau_{L,1}|^2 T_K/2}{(E-\epsilon+\Omega)^2 + (T_K/2)^2} \\
&= \frac{1}{2\pi} \left(\frac{|\tau_{L,1}|}{\tau_0}\right)^2 T_K \left[ \arctan\left(\frac{2\epsilon}{T_K}\right) + \frac{1}{2}\arctan\left(\frac{2\Omega - 2\epsilon}{T_K}\right) - \frac{1}{2}\arctan\left(\frac{2\Omega + 2\epsilon}{T_K}\right)\right] \\
&\overset{\Omega \gg \epsilon}{=} \frac{1}{2\pi} \left(\frac{|\tau_{L,1}|}{\tau_0}\right)^2 T_K \arctan\left(\frac{2\epsilon}{T_K}\right)
\end{aligned}
\tag{S3}
$$

for $\epsilon \ll \Omega \ll D$ and where $T_K = 4\tau_0^2/D$. An expansion in the limit of small driving frequency $\Omega$ yields the leading contribution $\sim \epsilon\,\Omega^2$. This renders an adiabatic approach insufficient to describe the pumped charge and thus underlines the relevance of the non-adiabatic physics at play. In the opposite limit of large driving frequency $\Omega \gg \epsilon$, plugging in the renormalized values for $\tau_0$ and $\tau_{L,1}$ from Eqs. (S1) and (S2) where only the second depends on $\Omega$ one recovers the power-law $J_{\mathrm{L},k=0} \sim \Omega^{-2U}$ mentioned in the main text (compare also Fig. 3 of the main text). Note that Eq. (S3) has the structure of a finite bias current through a non-interacting resonant level with bias voltage replaced by the onsite energy $\epsilon$.



\end{document}